\documentclass[12pt]{iopart}
\def\lsim{\mathrel{\rlap{
\lower4pt\hbox{\hskip-3pt$\sim$}}
    \raise1pt\hbox{$<$}}}     
\def\gsim{\mathrel{\rlap{
\lower4pt\hbox{\hskip-3pt$\sim$}}
    \raise1pt\hbox{$>$}}}     

\usepackage{graphicx}
\begin{document}
\title{Chemical Equilibrium in Heavy Ion Collisions: Rapidity Dependence.}
\author{F.~Becattini$^1$, J.~Cleymans$^2$ },
\address{$^1$ 1, Universit\`a di Firenze and INFN Sezione di Firenze \\
Largo E. Fermi 2, I-50125, Florence, Italy.\\
$^2$,UCT-CERN Research Centre and Department of Physics, 
University of Cape Town, ZA-7701 Rondebosch, South Africa}
\ead{becattini@fi.infn.it~~Jean.Cleymans@uct.ac.za}
\begin{abstract}
Particle yields in heavy ion collisions 
show an overwhelming evidence for
chemical or relative chemical equilibrium at all beam energies. 
The rapidity dependence of the 
thermal parameters $T$ and $\mu_B$ can now be determined over a wide range of
rapidities and show a systematic behavior towards an increase in $\mu_B$
away from mid-rapidity.
\end{abstract}
\section{Introduction}
\begin{figure}[thb]
\centerline{
\includegraphics[width=60mm,clip]{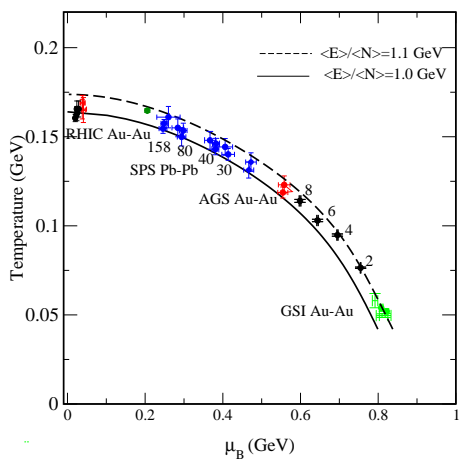}}
\caption{ Temperature vs. $\mu_B$ us determined from heavy ion 
collisions at different beam energies. The lower AGS points are 
based on a preliminary analysis of $4\pi$ data.}
 \label{eovern}

\end{figure}

Over the past decade, the analysis of particle 
multiplicities in heavy ion collions has shown overwhelming
evidence for chemical equilibrium in the final state except for
particles carrying strangeness which are mildly suppressed; however, their 
relative yields fulfill statistical equilibrium.
A summary  as of 2006, combining the results from many
different groups~\cite{comparison},  is shown in Fig. \ref{eovern}.
Except for particle multiplicities at RHIC energies, all data
in Fig.~\ref{eovern} use integrated particle yields, the
very systematic change of thermal parameters over the full range 
of beam energies is one of the most impressive features 
of relativistic ion collisions to date.
It is now possible to use the thermal model to make solid predictions for
particle multiplicities at LHC energies~\cite{kraus} and to determine
which beam energy will lead to the highest baryon density at 
freeze-out~\cite{randrup}. 
With chemical equilibrium for integrated yields thus firmly 
established,  we
focus on other properties, in particular,
since the rapidity distributions of identified particles is  now  becoming 
available also at RHIC energies~\cite{BRAHMS-shanghai},  it is now 
possible to determine the rapidity dependence of thermal parameters.
A first analysis was done by Stiles and Murray~\cite{stiles} for the 
data obtained by the
BRAHMS collaboration
at 200 GeV~\cite{BRAHMS,BRAHMS-shanghai}. This shows a clear dependence of 
the baryon chemical potential on rapidity in particular due to the 
changing $\bar{p}/p$ ratio.
Subsequently it was shown by R\"ohrich~\cite{roehrich} that 
the particle ratios at 
large rapidities  are consistent with those 
measured at the SPS energies. 
This opens the possibility to compare
measurements for e.g. the $K^+/\pi^+$ ratio at high rapidities and check them
with the corresponding values measured in the energy scan at the SPS, thus 
complementing the rapid variation of this ratio as a function of
beam energy. A further analysis of the rapidity dependence was recently done 
in Ref.~\cite{broniowski}. 
A detailed report of our results will be presented elsewhere~\cite{beca_jc}.
\section{Rapidity Distribution}
The general procedure is as follows: the rapidity axis is populated
with fireballs following a gaussian distribution function given by
$\rho(y_{FB})$ where $y_{FB}$ is the rapidity of the fireball. Particles 
will appear when the fireball freezes out and  will 
follow a thermal distribution  centered around the position
of the fireball 
\begin{equation}
\rho(y_{FB}) = \frac{1}{\sqrt{2}\pi\sigma} 
\exp\left( -\frac{y^2_{FB}}{2\sigma^2}\right) .
\label{eqrho}
\end{equation}

The momentum distribution of hadron $i$ is then calculated from 
the distribution of fireballs as given by Eq.~[\ref{eqrho}] along 
the rapidity axis as follows 
\begin{equation}
E_i\frac{d^3N_i}{d^3p} = \int_{-\infty}^{\infty} 
\rho\left(y_{FB}\right)E_i\frac{d^3N_1^i}{d^3p}(y-y_{FB})~dy_{FB}
\label{eqdist}
\end{equation}
where
$E_i\frac{d^3N_1^i}{d^3p}$ is the the distribution of hadrons 
from a single fireball.
The temperature $T$ and the baryon chemical potential $\mu_B$ will 
depend on the 
rapidity of the fireball and are not assumed to be constant.

An important parameter is the width of the distribution.
For the RHIC data at 200 GeV this  was determined from the 
  $\pi^+$'s as these are 
very sensitive to the value of $\sigma$ and  less to variations in
the baryon chemical potential. 
The width of the distribution 
$\sigma = 2.25$  is compatible with the values quoted by
the  BRAHMS collaborationi~\cite{BRAHMS},  e.g.
$\sigma_{\pi^+} = 2.25\pm 0.02$ 
and 
$\sigma_{\pi^-} = 2.29\pm 0.02$. 
The hadrons described by eq.~[\ref{eqdist}] are mainly resonances.
Only a fraction of these are stable under strong interactions.
The majority of them will decay
into stable hadrons at chemical freeze-out, hence the need to implement 
multi-particle decays.


\section{Freeze-Out Curve}
We assume that the temperature  $T$ and the chemical potential
$\mu_B$ are always related 
via the freeze-out curve as given in Fig.~[\ref{eovern}]; if the temperature 
varies along the rapidity axis, then also the
 chemical potential will vary. Thus a decrease in the temperature
of the fireball will be accompanied by an increase in the 
baryon chemical potential. 
In other words, we assue a universality of the chemical freeze-out condition.
This relationship between 
temperature and baryon chemical potential 
is  very reasonable since all particle abundancies measured so far 
follow it.
\begin{figure}[thb]
\centerline{
\includegraphics[width=60mm,clip]{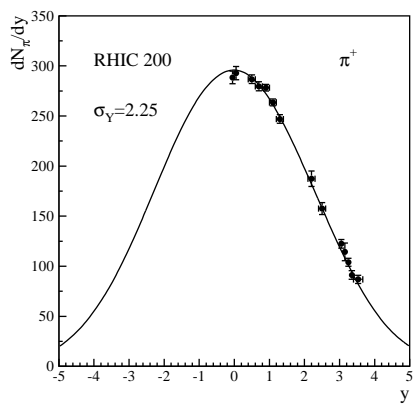}}
\caption{Fit to the pion distribution as measured by the BRAHMS
collaboration.}
 \label{fig:pions}
\end{figure}
Once the width of the distribution of fireballs has been fixed, we can go on with the
dependence of the baryon chemical potential on the rapidity of the fireball
(units are in GeV).
\begin{equation}
\mu_B = 0.026 + a~y_{FB}^2
\end{equation}

\begin{figure}[thb]
\centerline{
\includegraphics[width=60mm,clip]{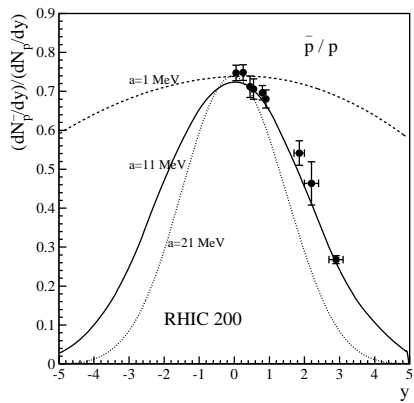}}
\caption{ The $\bar{p}/p$ ratio as a function of rapidity. The curves show the 
dependence on the baryon chemical potential on the rapidity as discussed in the text.}
 \label{fig:mub_y}
\end{figure}

The dependence of the $\bar{p}/p$ ratio on the 
parameter $a$ is shown in Figure [\ref{fig:mub_y}]. The best 
value is $a=0.012$ GeV.

\begin{figure}[thb]
\centerline{
\includegraphics[width=60mm,clip]{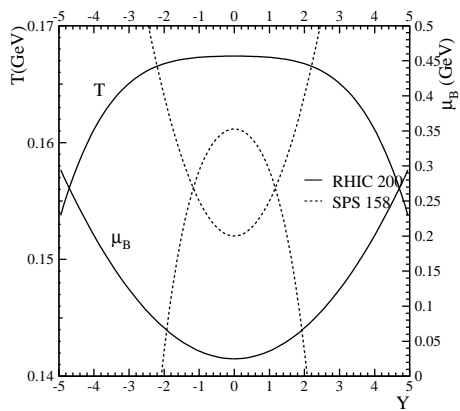}}
\caption{Values of the chemical freeze-out temperature and 
baryon chemical potential as a function of rapidity at the highest SPS (dotted lines) and RHIC (full lines) energies.}
 \label{fig:tmuy}
\end{figure}

The variation of the temperature along the rapidity axis is shown in 
Fig.~\ref{fig:tmuy}. The temperature is maximal at mid-rapidity
and gradually decreases towards higher (absolute) values of the 
rapidity. The baryon chemical potential follows a different pattern:
it has a minimum at mid-rapidity and increases quite 
substantially  towards higher values of the rapidity. 
Of particular interest are the 
largest values of the rapidity (say $y\approx 4$). 
It can be seen that there is a region of 
overlap with the data obtained at the NA49 experiment at CERN-SPS. 

\section{Summary}
Particle yields measured in heavy ion collisions show an overwhelming evidence for
chemical equilibrium at all beam energies. The rapidity dependence of the 
thermal parameters $T$ and $\mu_B$ can now be determined over a wide range of
rapidities and show a systematic behavior towards an increase in $\mu_B$
as the rapidity is increased.
\ack
This work was supported in part by the Scientific and
Technological Co-operation Programme between 
Italy and South Africa, project number 16. The help of 
D. Cebra with the preliminary 4$\pi$ data at the lower AGS beam energies is
gratefully acknowledged, R. Adams has helped in the 
analysis of these data in Fig.~[1] is acknowledged.

\end{document}